\documentclass[aps,preprint,prd,showpacs,nofootinbib]{revtex4}

\usepackage{amsmath,amssymb}
\usepackage{graphicx,subfigure}
\usepackage{color,multirow}
\usepackage[colorlinks,linkcolor=magenta,anchorcolor=cyan,citecolor=blue,plainpages=false]{hyperref}

\hypersetup{colorlinks=true,
    breaklinks=true,
    pdfstartview=Fit,
    linkcolor=blue,
    citecolor=blue,
    urlcolor=blue}

\def\be{\begin{equation}}
    \def\ee{\end{equation}}
\def\ba{\begin{eqnarray}}
    \def\ea{\end{eqnarray}}

\begin{document}

\title{Prepare inflationary universe via the Euclidean charged wormhole}

\author{Qing-Yu Lan$^{1} $\footnote{\href{lanqingyu19@mails.ucas.ac.cn}{lanqingyu19@mails.ucas.ac.cn}}}
\author{Yun-Song Piao$^{1,2,3,4} $ \footnote{\href{yspiao@ucas.ac.cn}{yspiao@ucas.ac.cn}}}

    \affiliation{$^1$ School of Physics Sciences, University of
        Chinese Academy of Sciences, Beijing 100049, China}

    \affiliation{$^2$ School of Fundamental Physics and Mathematical
        Sciences, Hangzhou Institute for Advanced Study, UCAS, Hangzhou
        310024, China}

    \affiliation{$^3$ International Center for Theoretical Physics
        Asia-Pacific, Beijing/Hangzhou, China}

    \affiliation{$^4$ Institute of Theoretical Physics, Chinese
        Academy of Sciences, P.O. Box 2735, Beijing 100190, China}

    \begin{abstract}

In this paper, we present a wavefunction of the universe, which
correspond to an Euclidean charged wineglass (half)-wormholes
semiclassically, as a possible creation for our inflationary
universe. We calculate the Euclidean action of the charged
wormhole, and find that the initial state of universe brought by
such an Euclidean charged wormhole can exhibit a high probability
weight for a long period of inflation. We compare our result with
that of axion wormholes, and evaluate the potential of other
corresponding Euclidean configurations as the pre-inflationary
initial states.

    \end{abstract}

    \maketitle

\tableofcontents

\section{INTRODUCTION}

The creation of the universe has been still a crucial issue in
cosmology and quantum gravity research in recent decades.
Inflation~\cite{Guth:1980zm,Linde:1981mu,Albrecht:1982wi,Starobinsky:1980te},
solving the issues of big-bang model, is the popular paradigm of
very early universe, and explains the existence of the primordial
perturbations in the early
universe~\cite{Mukhanov:1981xt,Starobinsky:1982ee,Guth:1982ec,Bardeen:1983qw,Hawking:1982cz}.

However, the initial conditions that started inflation in the very
early universe is not well understood. It can be expected that the
curvature of the universe and the matter densities arrives at the
Planck scale in the pre-inflationary epoch, rendering quantum
gravitational effects not negligible any more. This motivates the
consideration of an effective theory of quantum gravity (perhaps
not UV complete) to explore the issue.

Based on this consideration, the quantum state for the universe
emerging on a specific spacelike hypersurface $\Sigma(h_{ij},
\phi)$ can be written as $\Psi_{\Sigma}(h_{ij}, \phi)$.
This wavefunction should satisfy the Wheeler-DeWitt
equation~\cite{DeWitt:1967yk,Halliwell:1988ik}, which is a
second-order functional differential equation and requires the
boundary and initial conditions for a well-defined solution.
The Hartle-Hawking no-boundary wavefunction is a very intriguing
proposal for the wavefunction of the
universe~\cite{Hartle:1983ai,Hartle:2007gi,Blanco-Pillado:2011fcm},
which corresponds to a compact Euclidean semi-sphere
semiclassically. The probability weight of creating universe $P$
is set by the initial value of the inflation potential $V_{0}$,
the Hartle-Hawking no-boundary proposal suggests
\begin{equation}
    P(V_{0})=|\Psi|^{2}\simeq \exp\left(\frac{24\pi^{2}}{\kappa^{2}V_{0}}\right)\;.
\end{equation}

In this case, the universe with the least possible number of
inflationary e-folds is the most likely, since the weight is
larger for a smaller $V_{0}$, which implies that the
Hartle-Hawking no-boundary proposal seems to be incompatible with
a long-lasting period of inflation, see e.g. recent
Refs.~\cite{Maldacena:2024uhs,Lehners:2023yrj}. It is well known
that the scale-invariant primordial density perturbation predicted
by inflation has been confirmed by recent observations
\footnote{Though based on the $\Lambda$CDM model the Planck
collaboration showed the scalar spectral index is $n_s\approx
0.965$ \cite{Planck:2018jri}, $n_s=1$ ($n_s-1\sim {\cal O}
(0.001)$) is favored,
e.g.\cite{Ye:2020btb,Ye:2021nej,Jiang:2022uyg,Jiang:2022qlj,Ye:2022efx,Jiang:2023bsz,Peng:2023bik,Wang:2024dka,Wang:2024tjd},
when the pre-recombination resolution of Hubble tension is
considered.}, thus it is necessary to seek alternative initial
conditions for inflation.

In a theory of quantum gravity, the spacetime manifold should be
able to fluctuate. The fluctuations suggest that the spacetime
manifold may exhibit a foamy
structure~\cite{Wheeler:1957mu,Carlip:2022pyh}, leading to the
emergence of
wormholes~\cite{Hawking:1987mz,Hawking:1988ae,Coleman:1988cy,Giddings:1988cx,Giddings:1988wv}.
Previous studies indicate that the Euclidean (half)-wormholes can
lead to the creation of baby
universes~\cite{Giddings:1987cg,Betzios:2024oli,Jonas:2023ipa,Hebecker:2016dsw,Hebecker:2018ofv}.
The Euclidean wineglass-shaped axion wormholes correspond to the
emergence of expanding baby
universes~\cite{Betzios:2024oli,Jonas:2023ipa,Aguilar-Gutierrez:2023ril,Lavrelashvili:1988un},
as depicted in Fig. \ref{fig:wineglass wormhole}. The axion
wormholes were first found in the context of axion-scalar gravity
theory~\cite{Lavrelashvili:1988un,Rubakov:1988wx,Andriolo:2022rxc,Gutperle:2002km,Hertog:2017owm}.
Recently, they have been applied in Ref.~\cite{Betzios:2024oli} to
the very early universe as initial conditions for inflation.
The corresponding axion wormholes have an asymptotic anti-de
Sitter (AdS) boundary in the Euclidean past, which also have been
discussed earlier on wormhole applications in different
cosmologies~\cite{Maldacena:2004rf,Betzios:2019rds,Betzios:2021fnm,Antonini:2022blk,Antonini:2022ptt}.

In this paper, we present how the initial conditions of inflation
can be set by an Euclidean charged wormhole with the asymptotic
AdS boundary (see Fig.\ref{fig:wineglass wormhole}). The
corresponding wormholes are the Maxwell-scalar wormholes studied
widely in
Refs.~\cite{Marolf:2021kjc,Panyasiripan:2024iww,Hawking:1995ap,Dowker:1989uw,Kim:2001ri},
and are different from the axion wormholes in
Ref.~\cite{Betzios:2024oli}. In section \ref{model introduction}
and \ref{model wineglass}, we show the Euclidean charged wineglass
wormhole solution, and in section \ref{model on-shell action} we
further calculate the on-shell action. It is found that such
Euclidean charged wormholes, as the initial conditions for our
universe, can exhibit a higher probability weight for an universe
with sufficiently long inflationary period.

In section \ref{action comparison}, we further compare the
Euclidean action of our charged wormholes with that of axion
wormholes in Ref.~\cite{Betzios:2024oli}.
In section \ref{discussion}, we conclude our results and briefly
comment relevant issues. In Appendix \ref{app a8} and \ref{app
a2}, we discuss other Euclidean evolutions analogous to the method
used in section \ref{model wineglass} for the charged wormhole. In
Appendix \ref{app wormhole}, we discuss the Euclidean charged
wormhole with pure electro-magnetic field considered, this part of
discussion can be regarded as complementary to section~\ref{model
and action}.

\section{Euclidean charged wineglass wormhole}\label{model and action}
\subsection{Model}\label{model introduction}

Consider the simplest Euclidean spherically symmetric and isotropic metric
\begin{equation}\label{metric}
    ds^{2}=d \tau^{2}+a^{2}(\tau)d\Omega_{3}^{2}\;,
\end{equation}
\begin{equation}
    d\Omega_{3}^{2}=d\chi^{2}+\sin^{2}\chi(d\theta^{2}+\sin^{2}\theta d\phi^{2})\;.
\end{equation}
Analogous to that for the charged wormholes in
Ref.~\cite{Marolf:2021kjc}, we consider the Euclidean action as
\begin{equation}\label{action a4}
    S_{E}=\int d^{4}x\sqrt{g_{E}}(-\frac{1}{2\kappa}R+\frac{1}{2}\nabla^{\mu}\phi\nabla_{\mu}\phi+V(\phi)+F_{\mu\nu}F^{\mu\nu})\;.
\end{equation}
Here, $\kappa \equiv M_{pl}^{-2} = 8\pi G_{N}$. In corresponding
model, the Einstein gravity in four dimensions of spacetime is
minimally coupled to a scalar field, with a potential $V(\phi)$,
and an electro-magnetic field, which have been discussed in
Ref.~\cite{Marolf:2021kjc}, related to a certain Euclidean charged
wormholes with EAdS boundary conditions.
with the electromagnetic field strength tensor $F_{\mu\nu}$. Based
on the metric, we can calculate the Ricci curvature
\begin{equation}
    R_{\tau\tau}=-\frac{3a''}{a}\;,
\end{equation}
\begin{equation}
    R_{ij}=(-2-2a'^{2}-a a'')\delta_{ij}\;,
\end{equation}
\begin{equation}
    R=-6\left(\frac{-1}{a^{2}}+\frac{a'^{2}}{a^{2}}+\frac{-a a''}{a^{2}}\right)\;.
\end{equation}

Further, the equations of motion can be written as (corresponding
to the Einstein field equations, the scalar field equations, and
the electromagnetic field equations, respectively):
\begin{equation}
    R_{\mu\nu}-\frac{g_{\mu\nu}}{2}R=\kappa (T_{\mu\nu}^{Scalar}+T_{\mu\nu}^{EM})\;,
\end{equation}
\begin{equation}
    \phi''+3\frac{a'\phi'}{a}-\frac{dV}{d\phi}=0\;,
\end{equation}
\begin{equation}\label{EOMEM}
    \nabla_{\mu}F^{\mu\nu}=0\;,
\end{equation}
where the prime denotes derivatives with respect to the Euclidean
time, distinguishing it from the Lorentzian case. We first
consider the electromagnetic field equation of motion
(\ref{EOMEM}). Without loss of generality, we can express the
vector potential $\textbf{A}$ as \footnote{The reason for this
choice of $\textbf{A}$ will be explained in the Appendix \ref{app
wormhole}.}
\begin{equation}\label{vector potential}
    \textbf{A}=\left(0,-\frac{1}{2}\Phi(\tau),0,0\right)\;.
\end{equation}
Here, $\Phi(\tau)$ represents the electrostatic potential, and the
only non-zero component of the electromagnetic field tensor is
$F_{01} = -\frac{1}{2}\Phi'(\tau)$. Therefore, the equation of
motion (\ref{EOMEM}) can be simplified as
\begin{equation}
    \partial_{\mu}(\sqrt{g_{E}}F^{\mu\nu})=\partial_{0}(\sqrt{g_{E}}F^{01})=(-\frac{1}{2}a\Phi'(\tau))'=0\;.
\end{equation}
That is, $-\frac{1}{2}a\Phi'(\tau)$ should be related to an
integration constant $q$, and we get
\begin{equation}
    \Phi'(\tau)=\frac{q}{2\pi a}\;.
\end{equation}

Compared with the case of the Reissner-Nordström (RN) black hole,
it is clear that the meaning of this integration constant is the
charge. Thus we have
\begin{equation}
    F_{\mu\nu}F^{\mu\nu}=2F_{01}F^{01}=\frac{q^{2}}{8\pi^{2}}\frac{1}{a^{4}}=\frac{Q^{2}}{a^{4}}\;,
\end{equation}
where
\begin{equation}\label{EM charge}
    Q^{2}\equiv\frac{q^{2}}{8\pi^{2}}\;.
\end{equation}
Therefore, the action (\ref{action a4}) can be rewritten as
\begin{equation}\label{action a4}
    S_{E}=\int d^{4}x\sqrt{g_{E}}(-\frac{1}{2\kappa}R+\frac{1}{2}\nabla^{\mu}\phi\nabla_{\mu}\phi+V(\phi)+\frac{Q^{2}}{a^{4}})\;.
\end{equation}

\subsection{Wineglass wormhole}\label{model wineglass}
By substituting the Ricci curvature into the equations of motion, we can obtain the 00-component and the \(ij\)-component of the Einstein equations, as well as the equation of motion for the scalar field
\begin{equation}\label{eom einstein 00}
    \frac{a'^{2}}{a^{2}}-\frac{1}{a^{2}}+\frac{\kappa}{3}\left(V(\phi)-\frac{1}{2}\phi'^{2}\right)+\frac{\kappa Q^{2}}{3a^{4}}=0\;,
\end{equation}
\begin{equation}\label{eom einstein ij}
    \frac{2a''}{a} +\frac{a'^{2}}{a^{2}}-\frac{1}{a^{2}}+\kappa\left(V(\phi)+\frac{1}{2}\phi'^{2}\right)-\frac{\kappa Q^{2}}{3a^{4}}=0\;,
\end{equation}
\begin{equation}\label{eom scalar a4}
    \phi''+3\frac{a'\phi'}{a}-\frac{dV}{d\phi}=0\;.
\end{equation}
The scalar field equation (\ref{eom scalar a4}) can be viewed as a
``particle" $\phi$ moving in an effective potential $U_{eff} =
-V(\phi)$, with a friction term $3\frac{a'\phi'}{a}$ (if
$\frac{a'}{a} < 0$, it becomes an anti-friction term).

The Euclidean charged wineglass wormhole solution of interest
should have an asymptotically Euclidean AdS boundary as $\tau
\rightarrow \pm\infty$, i.e., $a(\tau) \sim \exp(H_{AdS}|\tau|)$.
In addition, we require that at $\tau = 0$, the Euclidean
evolution must convert to the Lorentzian evolution. To ensure that
this conversion corresponds to the creation of an inflationary
universe after switching to the Lorentzian time, at $\tau = 0$, we
further require\footnote{Generally for a wormhole, if the scale
factor $a(\tau)$ at $\tau = 0$ is finite, i.e. $a(0) = a_{0} \neq
0$, then for small $\tau$, we can expand $a$ as $a(\tau) = a_{0} +
\frac{1}{2}a''(0)\tau^{2} + O(\tau^{4})$. After analytically
continuing to Lorentzian evolution $t = -i\tau$, we obtain $a(t) =
a_{0} - \frac{1}{2}a''(0)t^{2} + O(t^{4})$. Hence, $a''(0) > 0$
corresponds to the creation of a collapsing universe, while
$a''(0) < 0$ corresponds to the creation of an inflationary
universe, where the wormhole at $\tau = 0$ is a local maximum. For
more discussion, see Ref.~\cite{Jonas:2023ipa}.}
\begin{equation}\label{initial conditions}
    a''(0)<0,\quad a'(0)=0,\quad a(0)=a_{max},\quad \phi'(0)=0\;.
\end{equation}
Substituting $a'(0)=0$ into equation (\ref{eom einstein 00}) and
letting $x \equiv a_{max}^{2}$, $V(\phi_{0})=V_{0}$, we obtain
\begin{equation}\label{A4 WORMHOLE}
    \tilde{V}_{0}x^{2}-x+\tilde{Q}^{2}=0\;,
\end{equation}
where we use the Planck units, so that
$\tilde{V}_{0}=\frac{\kappa}{3}V_{0}$ and
$\tilde{Q}^{2}=\frac{\kappa}{3}{Q}^{2}$. To ensure that such a
real solution for $a$ exists, the discriminant of equation
(\ref{A4 WORMHOLE}) must be positive, i.e.
\begin{equation}\label{judge a4}
    \Delta=1-4\tilde{V}_{0}\tilde{Q}^{2}>0\;,
\end{equation}
which suggests that the charge Q cannot be too large. In this
case, the two positive real solutions are
\begin{equation}\label{solution x}
    x_{1}=\frac{1+\sqrt{1-4\tilde{V}_{0}\tilde{Q}^{2}}}{2\tilde{V}_{0}}\;,\quad x_{2}=\frac{1-\sqrt{1-4\tilde{V}_{0}\tilde{Q}^{2}}}{2\tilde{V}_{0}}\;.
\end{equation}
Furthermore, we can observe the range of $x_1, x_2$
\begin{equation}
    x_{1}\in\left(\frac{1}{ 2\tilde{V}_{0}},\frac{1}{ \tilde{V}_{0}}\right),\quad x_{2}\in\left(0,\frac{1}{ 2\tilde{V}_{0}}\right)\;.
\end{equation}
This corresponds to the two solutions for the scale factor $a(\tau)$
\begin{equation}
    a_{1}\in \left(\sqrt{\frac{3}{2\kappa V(\phi_{0})}},\quad \sqrt{\frac{3}{\kappa V(\phi_{0})}}\right),\quad
    a_{2}\in \left(0,\quad \sqrt{\frac{3}{2\kappa V(\phi_{0})}}\right)\;.
\end{equation}
Substituting $a'(0)=0$ into equation (\ref{eom einstein ij}), we get
\begin{equation}
    a''=\frac{1}{2a}\left(1-\frac{2\kappa V(\phi_{0})}{3}a^{2}\right)\;.
\end{equation}

\begin{figure}
    \centering
    \includegraphics[width=0.8\linewidth]{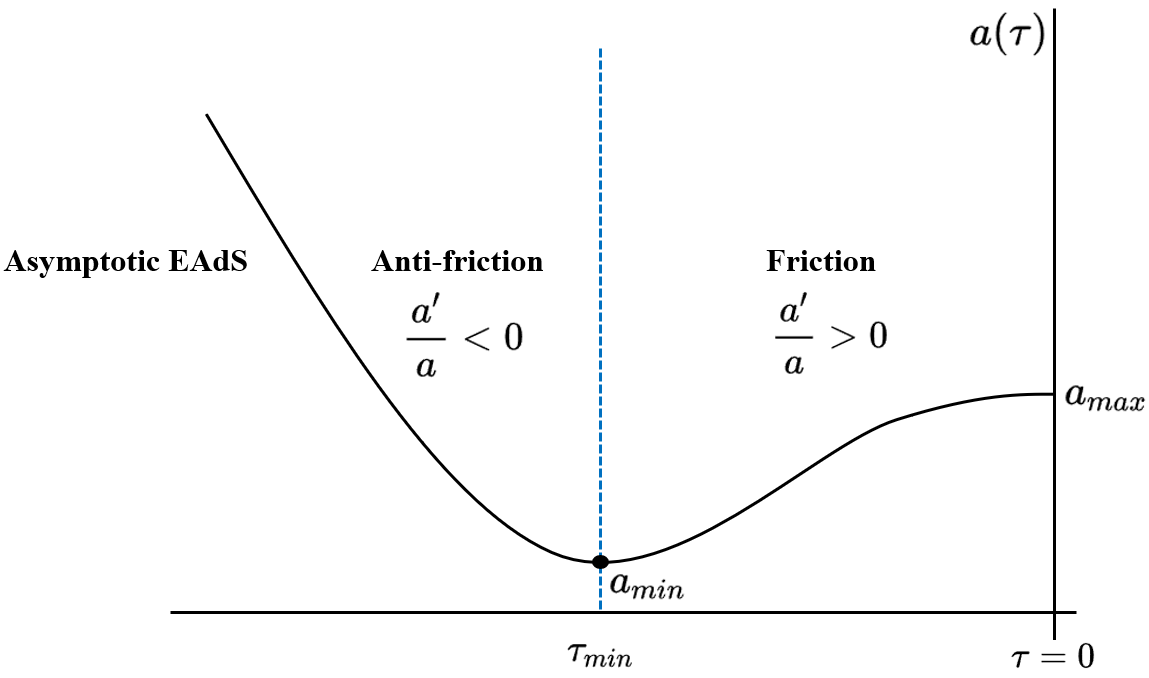}
\caption{The evolution of the scale factor $a(\tau)$ for a
Euclidean charged wineglass (half)-wormhole, here we defined
$a_1\equiv a_{max}\;, a_2\equiv a_{min}$. The dashed line marks
the point at which the scalar field $\phi(\tau)$ converts from the
anti-friction region to the friction region.}
    \label{fig:a(tau)}
\end{figure}

\begin{figure}
    \centering
    \includegraphics[width=0.55\linewidth]{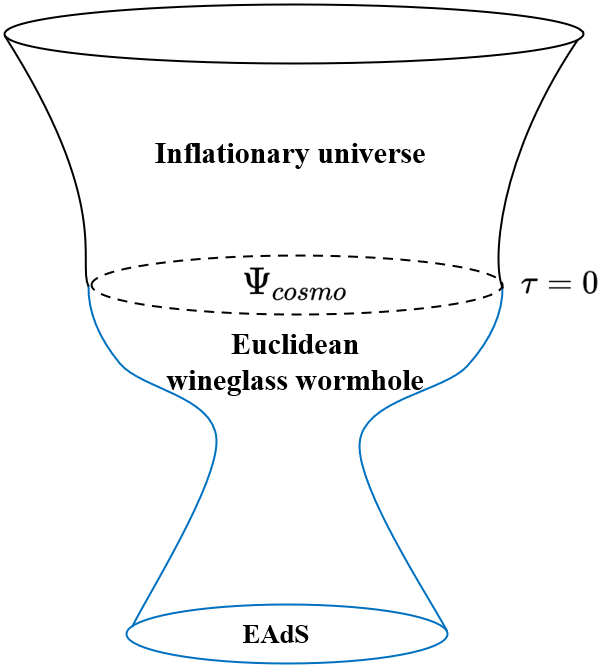}
\caption{The wavefunction for the universe calculated from
Euclidean path integral and the subsequent Lorentzian evolution.
The Euclidean evolution involves a charged wineglass
half-wormhole, resulting in an inflationary universe after
analytic continuation to Lorentzian signature at $\tau=0$
surface.}
    \label{fig:wineglass wormhole}
\end{figure}

It can be seen that $a_{1}>\sqrt{\frac{3}{2\kappa
V(\phi_{0})}}\;,\; a_{2}<\sqrt{\frac{3}{2\kappa V(\phi_{0})}}$. At
$a_{1}$ we have $a''<0$, corresponding to the maximum of the
wormhole; while at $a_{2}$ we have $a''>0$, corresponding to the
minimum of the wormhole. The evolution of the wormhole's scale
factor in the Euclidean time are depicted in Fig.
\ref{fig:a(tau)}, and considering the asymptotic Euclidean AdS
boundary, this corresponds to the Euclidean wineglass wormhole
illustrated in Fig. \ref{fig:wineglass wormhole}. Our results are
roughly consistent with the axion wormholes discussed in
Ref.~\cite{Betzios:2024oli}.

\subsection{The calculation of on-shell action}\label{model on-shell action}

Under the asymptotic Euclidean AdS boundary conditions, we can compute the on-shell action semi-classically by substituting the equations of motion into equation (\ref{action a4})
\begin{equation}
S_{E}^{on-shell}=4\pi^{2}\int
d\tau\left(\frac{Q^{2}}{a}-a^{3}V(\phi)\right)+S_{GHY}+S_{c.t.}\;.
\end{equation}
Here, $S_{GHY}$ is the Gibbons-Hawking-York boundary term and
$S_{c.t.}$ is the counterterm, which ensures the finiteness of the
on-shell action. Referring to Fig. \ref{fig:a(tau)}, we can divide
this integral into two parts, corresponding to the friction and
anti-friction regions. The first part includes the contribution
from the Euclidean AdS boundary term. The Euclidean AdS action
with an $S^{3}$ boundary has been calculated in earlier
studies~\cite{Jafferis:2011zi,Taylor:2016kic,Ghosh:2018qtg},
contributing a positive value to the action that does not depend
on the choice of the initial value $V(\phi_{0})$ of the inflation
potential. Therefore, we consider this part as a positive
constant. Next, we primarily focus on the integral for the second
part.

\subsubsection{Situation\;1:\; $a_{min}\ll a_{max}$}\label{situation 1}

In this situation, near $\tau_{min}$ in Fig. \ref{fig:a(tau)}, there is only a small region $\Delta \tau$ where $a' \simeq 0$, which we refer to as the "thin-wall"\footnote{This is analogous to the approximation method mentioned in Ref.~\cite{Coleman:1980aw}.}.  In this thin-wall region, we assume that the scalar field rapidly grows from $\phi_{\tau min}$ to near $\phi_{0}$, and the scale factor $a(\tau)$ is approximately a constant close to $a_{min}$. The action for the thin-wall region is
\begin{equation}
    S_{E}^{thin-wall}\simeq\frac{12\pi^{2}}{\kappa}\int_{thin}d\tau\left(\frac{\tilde{Q}^{2}}{a_{min}}-a_{min}^{3}\tilde{V}(\phi)\right)\simeq\frac{12\pi^{2}\tilde{Q}^{2}}{\kappa a_{min}}\Delta\tau_{thin}\;.
\end{equation}

This is similar to the boundary term that was previously neglected; it is also always positive and does not depend on $V(\phi_{0})$. Therefore, our main focus is on the contribution of the integral from the thick-wall region outside the thin-wall
\begin{equation}
  S_{E}^{thick-wall}\simeq\frac{12\pi^{2}}{\kappa}\int_{thick}d\tau\left(\frac{\tilde{Q}^{2}}{a}-a^{3}\tilde{V}(\phi_{0})\right) \;.
\end{equation}
Using equation (\ref{eom einstein 00}) to perform a change of
variables in the integral, we have
\begin{equation}\label{thick-wall a4}
     S_{E}^{thick-wall}\simeq\frac{12\pi^{2}}{\kappa}\int_{a_{min}}^{a_{max}}da\left(\frac{\tilde{Q}^{2}}{a\sqrt{1-a^{2}\tilde{V}(\phi_{0})-\frac{\tilde{Q}^{2}}{a^{2}}}}-\frac{a^{3}\tilde{V}(\phi_{0})}{\sqrt{1-a^{2}\tilde{V}(\phi_{0})-\frac{\tilde{Q}^{2}}{a^{2}}}}\right)\;.
\end{equation}

In the thick-wall region, the value of the scalar field is
approximately constant. Thus, $\tilde{V}$ in the above equation is
replaced by its value at $\phi_{0}$. Additionally, as noted from
equation (\ref{solution x}), $a_{min} \ll a_{max}$ is satisfied
only when $\tilde{Q} \rightarrow 0$,
Furthermore, since $a_{max} \simeq
\sqrt{\frac{1}{\tilde{V}(\phi_{0})}}$ when $a_{min} \ll a_{max}$,
after substituting the upper limit of the integral, equation
(\ref{thick-wall a4}) can be computed as
\begin{equation}\label{thick wall situation1}
    S_{E}^{thick-wall}\simeq\frac{12\pi^{2}\tilde{Q}^{2}}{\kappa}\left(-\log a_{min}\sqrt{\tilde{V}_{0}}+\log (1+\sqrt{1-a_{min}^{2}\tilde{V}_{0}})\right)-\frac{4\pi^{2}}{\kappa\tilde{V}_{0}}(1-a_{min}^{2}\tilde{V}_{0})^{\frac{3}{2}}\;.
\end{equation}
In the case where $a_{min} \ll a_{max}$ (i.e.,
$a_{min}\sqrt{\tilde{V}_{0}} \ll 1$), expanding the above equation
we get
\begin{equation}
     S_{E}^{thick-wall}\simeq\frac{12\pi^{2}\tilde{Q}^{2}}{\kappa}\left(-\log a_{min}\sqrt{\tilde{V}_{0}}+\log 2\right)-\frac{4\pi^{2}}{\kappa\tilde{V}_{0}}+ 
     O(a_{min}^{2}\tilde{V}_{0})\;.
\end{equation}
Neglecting the higher-order terms in
$a_{min}\sqrt{\tilde{V}_{0}}$, we observe that if $\tilde{Q} = 0$,
the action returns to the on-shell action in the Hartle-Hawking
no-boundary case. However, if $\tilde{Q}$ is small but not zero,
the action takes the form shown in the above equation. By taking
its derivative with respect to $\tilde{V}_{0}$, we get
\begin{equation}
\label{SE} \frac{\partial
S_{E}}{\partial\tilde{V}_{0}}=\frac{2\pi^{2}}{\kappa}\left(\frac{2-3\tilde{Q}^{2}\tilde{V}_{0}}{\tilde{V}_{0}^{2}}\right)\;.
\end{equation}

Since the corresponding state is a solution to the Wheeler-DeWitt
equation, the probability of universe creation, as a function of
$\tilde{V}_{0}$, is given by $P(\tilde{V}_{0}) = |\Psi|^{2} \simeq
e^{-S_{E}}$. It can be seen from the equation (\ref{SE}) that if
$\tilde{V}_{0}>\frac{2}{3\tilde{Q}^{2}}$, $S_{E}$ is smaller as
$\tilde{V}_{0}$ gets larger, which seems to imply that a universe
with long-lasting period of inflation has a higher probability of
being created, a behavior different from the Hartle-Hawking
no-boundary case.

However, it should be noted that the discriminant condition
(\ref{judge a4}) needs to be satisfied, i.e., $\tilde{V}_{0} <
\frac{1}{4\tilde{Q}^{2}} < \frac{2}{3\tilde{Q}^{2}}$, meaning that
the scenario where $\tilde{V}_{0} > \frac{2}{3\tilde{Q}^{2}}$ is
actually impossible. Therefore, although the action in section
\ref{situation 1} differs from the No-boundary case, the segment
of the solution that could resolve the sufficient inflation
problem is not accessible.

\subsubsection{Situation\;2:\;\;$a_{min}\simeq a_{max}$}\label{situation2}

In this situation, the region where $a'= 0$ is quite broad, and in
this thick-wall region, $a$ can be considered a constant. We can
replace $a$ with $\bar{a} = \frac{r}{\sqrt{\tilde{V}_{0}}}, (r
\sim O (1))$. At this point, the region where $a' \neq 0$ is very
narrow, corresponding to a thin-wall region. Since the potential
is approximately constant, as in section \ref{situation 1} , the
contribution of the thin-wall region to the integral should be
similar to the equation (\ref{thick wall situation1}). If we take
$a_{min} \simeq a_{max}$ in equation (\ref{thick wall
situation1}), the integral contribution becomes zero. Therefore,
the integral from the thin-wall region is independent of
$\tilde{V}_{0}$ and can be ignored in the first-order
approximation. Therefore, we only consider the contribution from
the thick-wall region, the Euclidean action is
\begin{equation}
    S_{E}\simeq\frac{12\pi^{2}}{\kappa}\int_{thick-wall}d\tau\left(\frac{\tilde{Q}^{2}}{\bar{a}}-\bar{a}^{3}\tilde{V}(\tilde{\phi})\right)\;.
\end{equation}

We use the scalar field equation of motion (\ref{eom scalar a4})
and integrating this equation over the Euclidean time from
$-\infty$ to $\tau$ to find
\begin{equation}\label{int eomscalar1}
    \frac{1}{6}\tilde{\phi}'^{2}-W_{friction}(\tau)=\tilde{V}(\tilde{\phi})-\tilde{V}_{\tau=-\infty}\;.
\end{equation}
Here, $\tilde{\phi}\equiv\phi/M_{Pl}$, and
$W_{friction}(\tau)=-\int_{-\infty}^{\tau}d\tilde{\tau}\frac{a'\tilde{\phi}'^{2}}{a}$
represents the "total work" done by the friction term as the
system evolves in the Euclidean time from $-\infty$ to $\tau$,
including the contribution of the friction region and
anti-friction region. According to the equation (\ref{int
eomscalar1}), around $\tilde{\phi} \simeq \tilde{\phi}_{\tau}$, we
have
\begin{equation}\label{int eomscalar2}
    \tilde{\phi}'=\sqrt{6\left(\tilde{V}(\tilde{\phi})-C\right)}\;.
\end{equation}
Here, $C \equiv -W_{f}$ is a constant, with its value ranging
between $V_{min} < C < V_{\tau}$. Using the equations (\ref{int
eomscalar1}) and (\ref{int eomscalar2}), we can write the
Euclidean action as
\begin{equation}
    S_{E}\simeq\frac{12\pi^{2}}{\kappa}\int_{\tilde{\phi}_{\tau min}}^{\tilde{\phi}_{0}}d\tilde{\phi}\left(\frac{\tilde{Q}^{2}/\bar{a}-\bar{a}^{3}\tilde{V}(\tilde{\phi})}{\sqrt{6(\tilde{V}(\tilde{\phi})-C)}}\right)\;.
\end{equation}

In this situation, we have the scalar field rolls from
$\tilde{\phi}_{\tau min}$ to near $\tilde{\phi}_{0}$ in the
thick-wall region where $a' \simeq 0$. In the vicinity of
$\tilde{\phi}_{0}$, the potential $\tilde{V}(\tilde{\phi})$ is
very close to $\tilde{V}_{0}$, and the potential can be expanded
around $\tilde{\phi}_{0}$ as $\tilde{V}(\tilde{\phi}) =
\tilde{V}_{0}(1 - \epsilon_{\tilde{V}}\tilde{\phi})$, where
$0<\epsilon_{\tilde{V}} \ll 1$ is related to the slow-roll
parameter. In this way, the action depends on $\tilde{V}_{0}$
through its dependence on $\tilde{\phi}_{0}$. The result of the
integral calculation reads
\begin{equation}\label{action sitution2 a4}
    S_{E}\simeq\sqrt{\frac{2}{3}}\frac{4\pi^{2}\sqrt{\tilde{V}_{0}-C}}{\kappa\bar{a}\epsilon_{\tilde{V}}\tilde{V}_{0}}\left(-3\tilde{Q}^{2}+\bar{a}^{4}(2C+\tilde{V}_{0})\right)\;.
\end{equation}
After substituting $\bar{a} = \frac{r}{\sqrt{\tilde{V}_{0}}}$, we
find that this action has a maximal value with respect to
$\tilde{V}_{0}$. This maximum of Euclidean action corresponds to a
certain potential $\tilde{V}_{*}$. If we can make the minimum of
the inflationary potential $\tilde{V}_{ms}$ \footnote{This value
differs from $\tilde{V}_{min}$, where the $\tilde{V}_{min}$ is the
overall negative minimum of the potential, and the
$\tilde{V}_{ms}$ is the minimum of the slow-roll inflationary
potential, see section \ref{supply potential} for detailed
discussion.} larger than $\tilde{V}_{*}$, then $S_{E}$ will be
smaller as $\tilde{V}_{0}$ gets larger. This leads to a higher
probability for the creation of a universe with long-lasting
period of inflation.

Next, we explain that this situation does not have the same
problem as section \ref{situation 1}. Substituting $\bar{a} =
\frac{r}{\sqrt{\tilde{V}_{0}}}$ and taking the derivative with
respect to $\tilde{V}_{0}$, we get
\begin{equation}\label{patial thick2}
    \frac{\partial S_{E}}{\partial \tilde{V}_{0}}\simeq-4\pi^{2}\frac{-10C^{2}r^{4}+2r^{4}\tilde{V}_{0}^{2}+C\tilde{V}_{0}(5r^{4}+3\tilde{Q}^{2}\tilde{V}_{0})}{\sqrt{6}r\kappa\epsilon_{\tilde{V}}\tilde{V}_{0}^{7/2}\sqrt{-C+\tilde{V}_{0}}}\;.
\end{equation}

To demonstrate that $S_{E}$ have access to the region where
$\tilde{V}_{0}$ is monotonically diminishing, we need to show that
the equation (\ref{patial thick2}) becomes negative as
$\tilde{V}_{0} \rightarrow \frac{1}{4\tilde{Q}^{2}}$ (where
$a_{min} \simeq a_{max}$). For this purpose, let's clarify the
other variables that appear in the equation.

Firstly, $\tilde{V}_{0}$ is the potential energy value at $\tau =
0$ in Planck units, i.e., the potential energy in the
pre-inflation epoch, and it is necessarily larger than zero. $C$
is the negative of the total work done during the interval $\tau
\in (-\infty, \tau_{min})$, with $\tilde{V}_{min} < C <
\tilde{V}_{\tau}$. Since the ``scalar particle" moves in the
effective potential $U_{eff} = -V(\phi)$, and both the initial and
final kinetic energies are zero, the total work from the friction
term is negative. Moreover, since Euclidean action depends on
$\tilde{V}_{0}$ through its dependence on $\tilde{\phi}_{0}$ and
$\tilde{\phi}$ is very close to $\tilde{\phi}_{0}$ at the end of
the thick-wall integral, we can approximate $C \simeq
\tilde{V}_{0}$, with the relation $C < \tilde{V}_{0}$ still holds.
Finally, $\epsilon_{\tilde{V}}$ is a positive coefficient and
$\epsilon_{\tilde{V}}\ll 1$, it is related to the slow-roll
parameter.

Further more, we provide a separate explanation for
$\bar{a}=\frac{r}{\sqrt{\tilde{V}_{0}}}$ in terms of $r$.
Considering that in section \ref{situation2}, $a_{min} \simeq
a_{max}$, and we know that in the limit $\tilde{V}_{0} \rightarrow
\frac{1}{4\tilde{Q}^{2}}$, $x_{1}, x_{2} \rightarrow
\frac{1}{2\tilde{V}_{0}}, \; x \equiv a^{2}$ from equation
(\ref{solution x}). We can also notice that the denominator in
equation (\ref{patial thick2}) is always positive, so we only need
to analyze the sign of the numerator. By replacing $\tilde{V}_{0}$
with $C$ in the numerator, and $\tilde{V}_{0}$ with
$\frac{1}{4\tilde{Q}^{2}}$, we obtain
\begin{equation}
    \frac{\partial S_{E}}{\partial \tilde{V}_{0}}\simeq\frac{3\left(r^{4}-\frac{1}{4}\right)}{4\tilde{Q}^{4}}\times\frac{\pi^{2}}{\sqrt{6}r\kappa\epsilon_{\tilde{V}}C^{7/2}\sqrt{-C+\tilde{V}_{0}}}\;.
\end{equation}

It can be divided into two parts, where the latter part is always
positive. Therefore, we only need to consider the sign of the
first part of the following equation. Note that although $r \simeq
\sqrt{\frac{1}{2}}$, $r$ never actually reaches
$\frac{1}{\sqrt{2}}$. This is because we must ensure that the
discriminant equation (\ref{judge a4}) in the wormhole solution
remains positive. For a more detailed expression, we can define
$\Delta \equiv k_1 ,\;0<k_1\ll 1$ in equation (\ref{judge a4}),
then
\begin{equation}\label{RE BAR A A4}
    \bar{a}= \frac{1}{2} \left( \frac{\sqrt{1 - \sqrt{k_1}}}{\sqrt{2\tilde{V}_{0}}} + \frac{\sqrt{1 + \sqrt{k_1}}}{\sqrt{2\tilde{V}_{0}}} \right)\;.
\end{equation}

It can be calculated that $\bar{a}$ asymptotically approaches
$\frac{1}{\sqrt{2}}$ from below. Thus under these conditions, it
is possible to reach the region where the action $S_E$ lowers with
$\tilde{V}_{0}$. This indicates that within the range allowed by
the discriminant, this situation does not exhibit the problem
present in section \ref{situation 1}.

\subsubsection{Supplementary discussion on the effective potential}\label{supply potential}

\begin{figure}
    \centering
    \includegraphics[width=1.0\linewidth]{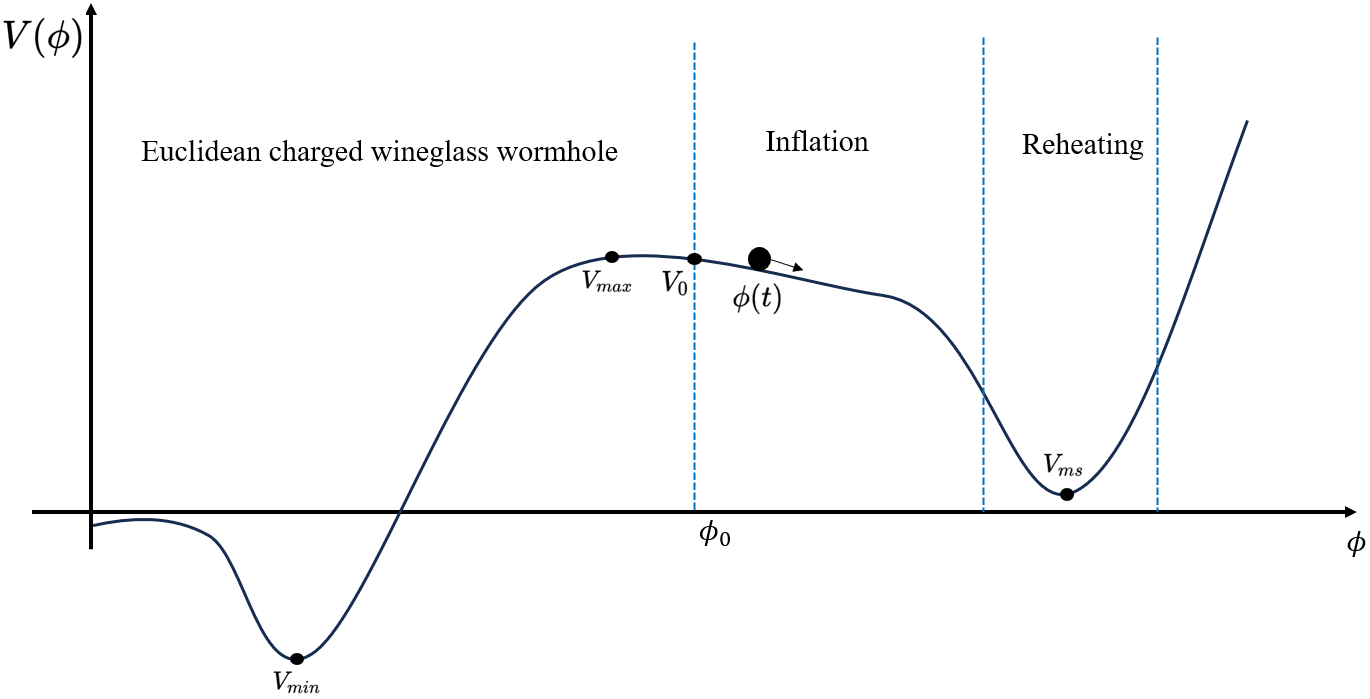}
\caption{The scalar potential $V(\phi)$. The left region
(wineglass wormhole) is Euclidean evolution region with
asymptotical EAdS boundary conditions. After the Euclidean
preparation, the universe evolves in the Lorentzian time $t$. We
sketch a slow-roll inflation potential with a reheating phase in
the Lorentzian evolution region.}
    \label{fig:potential V}
\end{figure}

The scalar potential we choose is shown in Fig. \ref{fig:potential
V}. Before $\phi_{0}$, it describes the Euclidean charged
wineglass wormhole. The Euclidean evolution, $\tau \in (-\infty,
0)$, prepares the initial state of the universe at $\phi_{0}$,
where it converts into Lorentzian evolution, followed by
inflation. As we mention in section \ref{model wineglass}, the
scalar field evolves in effective potential $-V(\phi)$, so the
Euclidean part of the potential should be fine-tuned to avoid
over/undergo behavior discussed in Ref.~\cite{Jonas:2023ipa}.
After $\phi_{0}$, we have connected a slow-roll inflationary
potential, which includes a reheating phase.

The chosen potential has the following characteristics: it has a
maximal value at $\phi=0$, $V_{\tau=-\infty}=V(\phi=0) \simeq 0$,
which corresponds to the asymptotic Euclidean AdS boundary.
Additionally, there is an overall minimal value $V_{min}$, a
positive maximal value $V_{max}$, and a positive metastable
minimum $V_{ms}$. These are the properties and parameters that the
potential model must have.

At $\phi_{0}$, the conversion into Lorentzian evolution occurs,
corresponding to a potential value $V_{0}$, which must satisfy $0
< V_{ms} \leq V_{0} < V_{max}$. In the region $\phi \in
(\phi_{max}, \phi_{0})$, the potential can be expanded as
$\tilde{V}(\tilde{\phi}) = \tilde{V}_{0} \left(1-
\epsilon_{\tilde{V}}(\tilde{\phi} - \tilde{\phi}_{0})\right)
\simeq \tilde{V}_{0} \left(1 - \epsilon_{\tilde{V}}
\tilde{\phi}\right), \; \epsilon_{\tilde{V}} \ll 1$. This is the
expansion we have chosen in section \ref{situation2}.

In the Lorentzian evolution region, the initial conditions
(\ref{initial conditions}) we set at $\tau=0$ are consistent with
an inflationary universe. The scalar field then evolves in a
slow-roll inflation potential as the one depicted in Fig.
\ref{fig:potential V}\footnote{The inflation potential $V(\phi) $
should satisfy the conditions of small slow-roll parameters in the
slow-roll approximation, i.e.,
$\epsilon_{V}=\frac{1}{16\pi\kappa}\left(\frac{V_{\phi}}{V}\right)^{2}\ll1,\;\eta_{V}=\frac{1}{8\pi\kappa}\left(\frac{V_{\phi\phi}}{V}\right)\ll1.\;$}.
We should also notice that the inflation is required to last
$N_{e-folds}\simeq O(60)$, where
\begin{equation}\label{efolds}
    N_{e-folds}\simeq -\kappa\int_{\phi_{0}}^{\phi_{e}}\frac{V}{V_{\phi}}d\phi\;.
\end{equation}
We integrate $\phi$ from $\phi_{0}$ to $\phi_{e}$ where the
inflation ends. The requirement for $N_{e-folds}$ can be satisfied
in various slow-roll models mentioned in
Refs.~\cite{Boubekeur:2005zm,Baumann:2014nda}.

\section{Compared with Euclidean axion wineglass wormhole}\label{action comparison}

In Ref.~\cite{Betzios:2024oli}, an Euclidean axion wineglass
wormhole model similar to the one discussed in section \ref{model
and action} is proposed. It also prepares the initial state of the
universe through the Euclidean evolution and addresses the problem
of sufficient inflation in the no-boundary proposal. However,
since our model encounters issues in the case where $a_{min} \ll
a_{max}$ and does not fully resolve the sufficient inflation
problem, we will focus the comparison in the situation where
$a_{min} \simeq a_{max}$.

Their and our models have consistent requirements for the scalar
potential and parameter selection. We assume that both evolve
under the same scalar potential as shown in Fig.
\ref{fig:potential V}. When $a_{min} \simeq a_{max}$, we define
the action of the axion wineglass wormhole as $S_{E}^{(1)}$, and
the action of the charged wineglass wormhole as $S_{E}^{(2)}$,
with the definition:
\begin{equation}
    \Delta S_{E}=S_{E}^{(1)}-S_{E}^{(2)}
\end{equation}
If $\Delta S_{E} > 0$, it indicates that the Euclidean charged
wineglass wormhole has a smaller action and dominates. Conversely,
if $\Delta S_{E} < 0$, the axion wormhole dominates.

\subsection{Wineglass wormhole}

According to Refs.~\cite{Betzios:2024oli,Jonas:2023ipa}, it is
known that the scale factor during the Euclidean evolution are
same as in Fig. \ref{fig:a(tau)}, where $a(\tau)$ has a minimal
value and a local maximum. The two solutions are ($x \equiv
a^{2}$)
\begin{equation}\label{solution x a6}
    x_{1}=\frac{1}{3\tilde{V}_{0}}\left(1+2\cos \left(\frac{\theta}{3}\right)\right),\quad x_{2}=\frac{1}{3\tilde{V}_{0}}\left(1+2\cos \left(\frac{\theta-2\pi}{3}\right)\right)\;,
\end{equation}
where
\begin{equation}
    \cos \theta=1-\frac{27}{2}\tilde{Q}^{2}\tilde{V}_{0}^{2},\quad\theta\in(0,\pi )\;,
\end{equation}
\begin{equation}
    x_{1}\in\left(\frac{2}{3\tilde{V}_{0}},\frac{1}{\tilde{V}_{0}}\right),\quad x_{2}\in\left(0,\frac{2}{3\tilde{V}_{0}}\right)\;.
\end{equation}
The discriminant is given by
\begin{equation}
    \Delta=\tilde{Q}^{2}(4-27\tilde{Q}^{2}\tilde{V}_{0}^{2})>0\;.
\end{equation}
In the situation where $a_{min} \simeq a_{max}$, the on-shell
Euclidean action is
\begin{equation}\label{action situation2 a6}
    S_{E}^{(1)}\simeq\sqrt{\frac{2}{3}}\frac{4\pi^{2}\sqrt{\tilde{V}_{0}-C^{(1)}}}{\kappa\bar{a}^{3}\epsilon_{\tilde{V}}\tilde{V}_{0}}\left(-6\tilde{Q}_{(1)}^{2}+\bar{a}^{6}(2C^{(1)}+\tilde{V}_{0})\right)\;.
\end{equation}

In this case, we also have $\bar{a} =
\frac{r}{\sqrt{\tilde{V}_{0}}}$, but the value of $r$ needs to be
re-estimated. However, if $a_{min} \simeq a_{max}$, the equation
(\ref{solution x a6}) requires $\theta \rightarrow \pi$, which
means $\tilde{V}_{0}^{2} \rightarrow \frac{4}{27\tilde{Q}^{2}}$,
and in this case, $r \rightarrow \sqrt{\frac{2}{3}}$. For a more
detailed expression like equation (\ref{RE BAR A A4}), we set
\begin{equation}
    4-27\tilde{Q}^{2}\tilde{V}_{0}^{2}\equiv k_{2},\quad 0<k_{2}\ll 1\;.
\end{equation}
Now we can find
\begin{equation}\label{RE BARA A6}
    \theta\simeq\pi-\sqrt{k_{2}},\quad \bar{a}=\frac{\sqrt{\frac{1 - 2 \sin\left(\frac{1}{6} \left(2 \sqrt{k_2} - \pi\right)\right)}{\tilde{V}_{0}}} + \sqrt{\frac{1 + 2 \sin\left(\frac{1}{6} \left(2 \sqrt{k_2} + \pi\right)\right)}{\tilde{V}_{0}}}}{2 \sqrt{3}}\;.
\end{equation}

Additionally, we need to clarify the constant $C^{(1)}$ in this
situation. $C^{(1)}$ represents the negative of the total work
done by the ``scalar particle" during the Euclidean evolution
$(-\infty, \tau)$ in the axion wineglass wormhole. At the end of
the thick wall integration where $a_{min} \simeq a_{max}$, $\phi$
is already very close to $\phi_{0}$, so in fact, $C^{(1)} \simeq
\tilde{V}_{0}$.

Here, we are regarding $\Delta S_{E}$ as a function of
$\tilde{V}_{0}$ and comparing the action values of both cases
under the same $\tilde{V}_{0}$, we can assume that the constant
$C$ is the same in both cases. Therefore, we have
\begin{equation}
    C^{(1)}\simeq C^{(2)}=C\;,
\end{equation}
always holds.

\subsection{Compare with our charged actions}

By combining equation (\ref{action sitution2 a4}) with equation
(\ref{action situation2 a6}) and substituting the corresponding
parameters, we note that $Q$ and $r$ differ in these two cases. In
$S_{E}^{(1)}$, we substitute
$\tilde{Q}_{(1)}^{2}=\frac{4-k_2}{27\tilde{V}_{0}^{2}}$ and
replace $\bar{a}$ using Equation (\ref{RE BARA A6}), while in
$S_{E}^{(2)}$, we substitute
$\tilde{Q}_{(2)}^{2}=\frac{1-k1}{4\tilde{V}_{0}}$ and replace
$\bar{a}$ using Equation (\ref{RE BAR A A4}). We compare the
magnitude of their actions

\begin{equation}
    \Delta S_{E}=S_{E}^{(1)}-S_{E}^{(2)}\;.
\end{equation}

\begin{figure}
    \centering
    \includegraphics[width=0.7\linewidth]{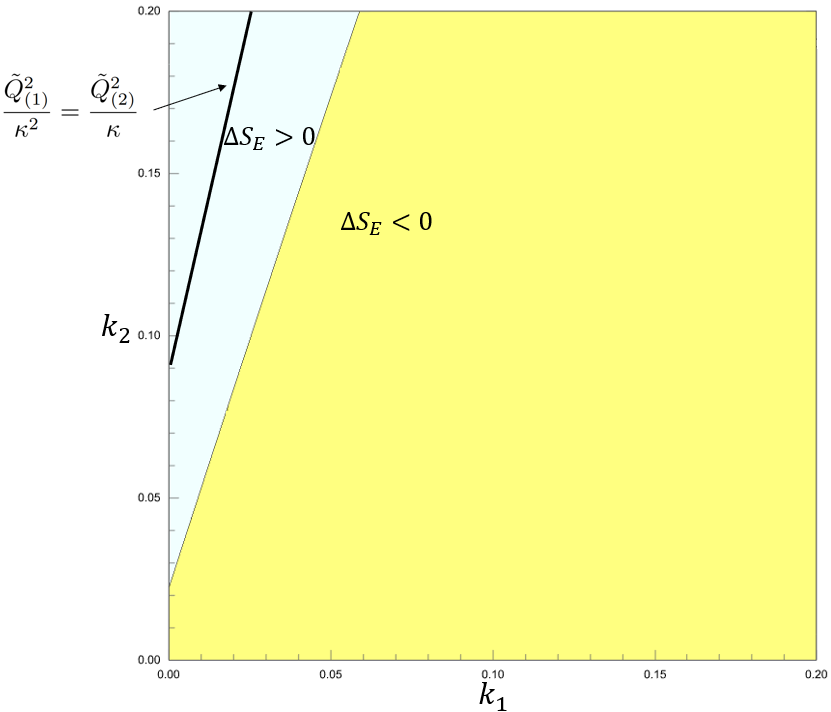}
\caption{Comparison of Euclidean actions $\Delta S_{E}$ as a
function of $k_1$ and $k_2$. The black line in figure represents
the relationship that $k_1$ and $k_2$ must satisfy when the
dimensionless charge of the axion wormhole is the same as that of
the charged wormhole. The parameters used are
$\kappa\tilde{V}_{0}=0.58,\;C=0.99\tilde{V}_{0},\;\epsilon_{\tilde{V}}=0.1$.}
    \label{fig:SameQDeltase}
\end{figure}

The specific form of $\Delta S_{E}$ is quite complex. We assign a
specific value to $\kappa\tilde{V}_{0}$ and consider $\Delta
S_{E}$ as a function of $k_1$ and $k_2$. Since both $k_1$ and
$k_2$ are small quantities, it is reasonable to choose their range
as $(0, 0.2)$. Thus we can get Fig. \ref{fig:SameQDeltase}.

The yellow region represents the area where $\Delta S_{E} < 0$,
indicating the dominance of the axion wineglass wormhole. The blue
region satisfies $\Delta S_{E} > 0$, where the charged wineglass
wormhole dominates. It is also worth noting that such results are
independent of the value of $\kappa\tilde{V}_{0}$, since with
other parameter selections of $\kappa\tilde{V}_{0}$ we can obtain
similar results in the region where both $k_1$ and $k_2$ are
small. There will always exist two regions: one dominated by the
charged wineglass wormhole and the other by the axion wineglass
wormhole.

We further compare the magnitudes of the two actions under the
same dimensionless wormhole charge and the same initial condition
of the inflation field, with $\kappa\tilde{V}_{0} = 0.58$.
However, it should be noted that the dimensions of the electric
charge and the axion charge are different. Therefore, we first
need to nondimensionalize them before making the
comparison\footnote{We can use the dimensionless nature of the
Euclidean action (\ref{action sitution2 a4}) (\ref{action situation2 a6}) to nondimensionalize the wormhole charge.}, that is
\begin{equation}\label{same dimensionless Q}
    \frac{\tilde{Q}_{(1)}^{2}}{\kappa^{2}}=\frac{\tilde{Q}_{(2)}^{2}}{\kappa}\;.
\end{equation}
At this point, we take $\kappa\tilde{V}_{0} = 0.58$, the equation
(\ref{same dimensionless Q}) gives us a linear relation between
$k_1$ and $k_2$, which is represented by the black line in Fig.
\ref{fig:SameQDeltase}.
We find that, under the same dimensionless wormhole charge and the
same initial value of the inflation field $\kappa\tilde{V}_{0}$,
there is always $\Delta S_{E} > 0$. Therefore, since the
probability of the creation of universe is given by
$P(\tilde{V}_{0}) = |\Psi|^{2} \simeq e^{-S_{E}}$, in this case
the Euclidean charged wineglass wormhole always dominates.

We can also compare the Euclidean action of the charged wormhole
with that of the no-boundary proposal. In this case, it should be
noted that this comparison is valid only under the conditions
specified in section \ref{situation 1} since the evolution of
scale factor in no-boundary proposal satisfies $a_{min}\ll
a_{max}$. We find that
\begin{equation}
    \Delta S_{E}=S_{E}^{charged}-S_{E}^{no-boundary}\simeq-\frac{12\pi^{2}\tilde{Q}^{2}}{\kappa}\log a_{min}\sqrt{\tilde{V}_{0}}\gg 1\;.
\end{equation}
This indicates that the no-boundary Euclidean evolution always
dominates the creation of inflationary universes, similar to the
case in Refs.~\cite{Betzios:2024oli,Friedrich:2024aad}.

\section{Conclusion and outlook}\label{discussion}

In this paper, we propose a different initial condition for
inflation, which correspond to an Euclidean charged wineglass
(half)-wormholes semiclassically. We calculate the Euclidean
action of the charged wormhole, and find that the initial state of
universe brought by such Euclidean charged wormholes can exhibit a
high probability weight for a long period of inflation.
We further compare its Euclidean action with that of the axion
wormhole in Ref.~\cite{Betzios:2024oli}, and observe that when
their dimensionless charge are equal, the probability of charged
wormhole responsible for inflation is always larger, see Fig.
\ref{fig:SameQDeltase}.

In our proposal, the wavefunction of the universe should be the
solution of the Wheeler-DeWitt equation~\cite{DeWitt:1967yk}, its
potential includes a similar slow-roll approximation part (see
Fig. \ref{fig:potential V}.).
Thus our chosen initial conditions can also naturally predict the
correct primordial perturbation spectrum. The slow-roll inflation
might happen in a landscape with lots of dS and AdS vacua,
e.g.\cite{Li:2009me,Huang:2023chx,Huang:2023mwy}, in which
supermassive primordial black holes (a possible interpretation of
high-redshift massive galaxies observed by JWST,
e.g.\cite{Hai-LongHuang:2024kye,Hai-LongHuang:2024vvz,Hai-LongHuang:2024gtx})
can naturally come into being by the nucleation of supercritical
dS bubbles, thus it is interesting to explore how to apply our
proposal in such a landscape\footnote{It is possible that inflaton
consists of mutiple scalar fields, in this case the spectrum of
scalar perturbation is generally redder, e.g.\cite{Piao:2006nm}.
}.

Inspired by the case of the Euclidean charged wormhole, it will be
interesting to consider different Euclidean evolutions as initial
conditions for inflation. In Appendix \ref{app a8} and \ref{app
wormhole}, the initial states we get through Euclidean preparation
correspond to cosmologies inevitably crunch, and we get a
no-boundary-like initial state in Appendix \ref{app a2}. Both
cases do not seem to have a sufficient period of inflation.
However, it is also noted that in all cases considered, the
no-boundary proposal has always higher weight for the creation of
universes, see also recent Ref.~\cite{Friedrich:2024aad}.
Therefore, it would be clearly important and still left to find an
Euclidean evolution not only responsible for the initial state of
long-lasting inflationary period but also overwhelm the weight of
the Hartle-Hawking no-boundary state.

Another interesting idea is to consider the implication of the
AdS/CFT duality~\cite{Maldacena:1997re,Witten:1998qj} for our Euclidean charged wormhole proposal, since
there have been holographic considerations in no-boundary
proposal~\cite{Hertog:2011ky,Hartle:2012tv}.
It is significant to find an EFT associated with a holographic CFT
to describe our Euclidean
evolution~\cite{Antonini:2022blk,VanRaamsdonk:2024sdp}, since the
wineglass wormholes we consider have EAdS boundary conditions. It
is also interesting to consider whether the entanglement island
can emerge in such a gravitational prepared
state~\cite{Chen:2020tes,Fumagalli:2024msi} and the dS-like
inflationary universe in the subsequent Lorentzian evolution,
e.g.~\cite{Piao:2023vgm,Jiang:2024xnd,Chang:2023gkt}.


\section*{Acknowledgments}

This work is supported by NSFC (Grant No.12075246), National Key
Research and Development Program of China (Grant No.
2021YFC2203004), and the Fundamental Research Funds for the
Central Universities.

\appendix
\section{Euclidean action includes the term ${Q^{2}}/{a^{8}}$}\label{app a8}

In this section, we consider the Euclidean evolution action with
the term $\frac{Q^{2}}{a^{8}}$. The action for this Euclidean
evolution is
\begin{equation}\label{action a8}
    S_{E}=\int d^{4}x\sqrt{g_{E}}(-\frac{1}{2\kappa}R+\frac{1}{2}\nabla^{\mu}\phi\nabla_{\mu}\phi+V(\phi)+\frac{Q^{2}}{a^{8}})\;.
\end{equation}
The corresponding equation of motion is
\begin{equation}\label{eom einstein 00 a8}
    \frac{a'^{2}}{a^{2}}-\frac{1}{a^{2}}+\frac{\kappa}{3}\left(V(\Phi)-\frac{1}{2}\phi'^{2}\right)+\frac{\kappa Q^{2}}{3a^{8}}=0\;,
\end{equation}
\begin{equation}\label{eom einstein ij a8111}
    \frac{2a''}{a} +\frac{a'^{2}}{a^{2}}-\frac{1}{a^{2}}+\kappa\left(V(\Phi)+\frac{1}{2}\phi'^{2}\right)-\frac{5\kappa Q^{2}}{3a^{8}}=0\;,
\end{equation}
\begin{equation}\label{eom scalar}
    \phi''+3\frac{a'\phi'}{a}-\frac{dV}{d\phi}=0\;.
\end{equation}
We calculate the wormhole solution similarly to the previous case.
In equation (\ref{eom einstein 00 a8}), by setting $a' = 0$,
$\phi' = 0$, $\tau = 0$, and $x \equiv a^{2}$, we get
\begin{equation}\label{a8 x4}
    \tilde{V}_{0}x^{4}-x^{3}+\tilde{Q}^{2}=0 \;.
\end{equation}

This is a quartic equation, and the solution is much more complex
compared to the Euclidean charged wormhole and the axion wormhole
solution in Ref.~\cite{Betzios:2024oli}. Assuming there are four
real solutions, we first consider the properties of these
solutions. Only the solutions with $a'' < 0$ correspond to the
creation of an inflationary universe. From equation (\ref{eom
einstein ij a8111}), we know that
\begin{equation}
    a''=\frac{3-4a^{2}\tilde{V}_{0}}{a}\;.
\end{equation}
Therefore, only the solutions that satisfy $a >
\sqrt{\frac{3}{4\tilde{V}_{0}}}$ are the ones we are looking for,
which can gives rise to an expanding universe. We define
\begin{equation}
E = \frac{4 \left(\frac{2}{3}\right)^{1/3} \tilde{Q}^2}{\left(9 \tilde{Q}^2 + \sqrt{3} \sqrt{27 \tilde{Q}^4 - 256 \tilde{Q}^6 \tilde{V}_{0}^3}\right)^{1/3}} + \frac{\left(9 \tilde{Q}^2 + \sqrt{3} \sqrt{27 \tilde{Q}^4 - 256 \tilde{Q}^6 \tilde{V}_{0}^3}\right)^{1/3}}{2^{1/3} 3^{2/3} \tilde{V}_{0}}\;.
\end{equation}

To ensure that the equation has real roots ($\tilde{Q}$ cannot be
too large), $E$ must satisfy $E > 0$ and $E \ll 1$. The largest
solution of equation (\ref{a8 x4}) can be written as
\begin{equation}
x_{max} = \frac{1}{2} \sqrt{E + \frac{1}{4 \tilde{V}_{0}^2}} + \frac{1}{4 \tilde{V}_{0}} + \frac{1}{2} \sqrt{\frac{1}{4 \sqrt{E + \frac{1}{4 \tilde{V}_{0}^2}} \tilde{V}_{0}^3} + \frac{1}{2 \tilde{V}_{0}^2} - E}\;.
\end{equation}
Expanding the solution for the largest root in the case where
$E\ll 1$, we get
\begin{equation}
    x_{max}\simeq\frac{3}{4\tilde{V}_{0}}-\frac{\tilde{V}_{0}E}{2}\;,\quad a_{max}<\sqrt{\frac{3}{4\tilde{V}_{0}}}\;.
\end{equation}
If $a_{max}<\sqrt{\frac{3}{4\tilde{V}_{0}}}$, then $a''>0$ always
holds, and all the solutions of equation (\ref{a8 x4}) correspond
to the creation of a collapsing universe. This is not the suitable
initial state for inflation we are looking for.

\section{Euclidean action includes the term ${Q^{2}}/{a^{2}}$}\label{app a2}

In this section, we consider the Euclidean action with a
$\frac{Q^{2}}{a^{2}}$ term. The result is similar to the
No-boundary case, where only the coupling between gravity and the
scalar field is considered. The form of the action is
\begin{equation}\label{action a2}
    S_{E}=\int d^{4}x\sqrt{g_{E}}(-\frac{1}{2\kappa}R+\frac{1}{2}\nabla^{\mu}\phi\nabla_{\mu}\phi+V(\phi)+\frac{Q^{2}}{a^{2}})\;.
\end{equation}
The corresponding equation of motion is
\begin{equation}\label{eom einstein 00 a2}
    \frac{a'^{2}}{a^{2}}-\frac{1}{a^{2}}+\frac{\kappa}{3}\left(V(\Phi)-\frac{1}{2}\phi'^{2}\right)+\frac{\kappa Q^{2}}{3a^{2}}=0\;,
\end{equation}
\begin{equation}\label{eom einstein ij a2}
    \frac{2a''}{a} +\frac{a'^{2}}{a^{2}}-\frac{1}{a^{2}}+\kappa\left(V(\Phi)+\frac{1}{2}\phi'^{2}\right)+\frac{\kappa Q^{2}}{3a^{2}}=0\;,
\end{equation}
\begin{equation}\label{eom scalar}
    \phi''+3\frac{a'\phi'}{a}-\frac{dV}{d\phi}=0\;.
\end{equation}
By setting $a'=0$, $\phi'=0$, and $\tau=0$ in equation (\ref{eom
einstein 00 a2}), we get
\begin{equation}
    a^{2}=\frac{1-\tilde{Q}^{2}}{\tilde{V}_{0}}\;.
\end{equation}
When $\tilde{Q}^{2} \ll 1$, $a$ has only one solution. If we set
$\tilde{Q} = 0$, it returns to the no-boundary proposal. We
continue to use equation (\ref{eom einstein ij a2}) to analyze the
extremal properties of this solution, and we find
\begin{equation}
    a''=\frac{a}{2}\left(\frac{1-\tilde{Q}^{2}}{a^{2}}-3\tilde{V}_{0}\right)=-a\tilde{V}_{0}<0\;.
\end{equation}
Therefore, this solution corresponds to the maximal value of the
scale factor, which can give rise to the creation of an
inflationary universe. We further analyze the on-shell action;
substituting the equations of motion, we obtain
\begin{equation}
    S_{E}^{on-shell}=\frac{12\pi^{2}}{\kappa}\int d\tau\sqrt{g_{E}}\left(-\tilde{V}(\phi)\right)\;.
\end{equation}

This is precisely the on-shell action for the minimal coupling
between gravity and the scalar field under the no-boundary
condition as proposed by Hartle-Hawking. In this case, just like
in the no-boundary proposal, there is still the issue of
insufficient inflation. This is not the initial state for
inflation we are looking for either.

\section{Charged Euclidean wormhole with EAdS boundary}\label{app wormhole}
\subsection{The choose of charged Euclidean wormhole action}

This subsection provides an explanation for the choice of the
electromagnetic vector potential. This part primarily refers the
discussion in Ref.~\cite{Marolf:2021kjc}. The Euclidean action
considered is
\begin{equation}
    S_{E} = - \int_{\mathcal{M}} d^4x \sqrt{g_{E}} \left( \frac{1}{2\kappa}R + \frac{6}{L^2} - \sum_{I=1}^{3} F_{\mu\nu}^{(I)} F^{(I) \mu\nu} \right) - 2 \int_{\partial \mathcal{M}} d^3x \sqrt{h} K + S_{GHY}\;.
\end{equation}

The above equation considers three independent Maxwell fields,
which is due to the fact that the metric (\ref{metric}) has a
3-sphere symmetry (i.e., $SO(4)$ symmetry). Introducing a single
Maxwell field would break this symmetry. To obtain the specific
form of the vector potential, we rewrite the 3-sphere metric
$d\Omega^{3}$ in terms of Euler angles in a more symmetric form
\begin{equation}
    d\Omega^2_3 = \frac{1}{4} \left( \hat{\sigma}_1^2 + \hat{\sigma}_2^2 + \hat{\sigma}_3^2 \right)\;,
\end{equation}
where
\begin{equation}
    \hat{\sigma}_1 = - \sin \psi \, d\theta + \cos \psi \, \sin \theta \, d\phi \;,
\end{equation}
\begin{equation}
    \hat{\sigma}_2 = \cos \psi \, d\theta + \sin \psi \, \sin \theta \, d\phi \;,
\end{equation}
\begin{equation}
    \hat{\sigma}_3 = d\psi + \cos \theta \, d\phi\;.
\end{equation}
Based on this, we choose the vector potential as
\begin{equation}
    \textbf{A}^{(I)} =  \frac{\hat{\sigma}_I}{2} \, \Phi(\tau),  \quad I \in \{1, 2, 3\}\;.
\end{equation}

The form of the vector potential we chose in section \ref{model
introduction} (equation (\ref{vector potential})) is derived from
this. However, in section \ref{model introduction}, we considered
the situation where only one of the three electromagnetic fields
was taken into account. This is because the three electromagnetic
fields are completely independent of each other, each satisfying
its own equation of motion, and the analysis for each is identical
to the process described in section \ref{model introduction}.

To account for the contributions of all three electromagnetic
fields without breaking the symmetry, we simply need to redefine
$Q^{2}$ in equation (\ref{EM charge}) as
\begin{equation}
    Q^{2}\equiv\frac{1}{8\pi^{2}}\sum_{I=1}^{3}q_{I}^{2}\;,
\end{equation}
this does not affect the subsequent discussion.

\subsection{Wormhole}

In this subsection, we consider the Euclidean action with only
gravitational and electromagnetic field terms, with an EAdS
boundary. The form of the action is\footnote{In the case of the
charged wormhole discussed earlier, there is also an asymptotic
Euclidean AdS boundary. However, the terms related to AdS in
equation (\ref{action EADS PURE EM}) can be absorbed into the
scalar potential.}
\begin{equation}\label{action EADS PURE EM}
    S_{E}=\int d^{4}x\sqrt{g_{E}}(-\frac{1}{2\kappa}R+\frac{6}{L^{2}}+F_{\mu\nu}F^{\mu\nu})
\end{equation}\;
Where $L=1$ is the AdS radius, the corresponding equation of motion is
\begin{equation}\label{eom einstein 00 em}
    \frac{a'^{2}}{a^{2}}-\frac{1}{a^{2}}+\frac{\kappa Q^{2}}{3a^{4}}-\frac{1}{L^{2}}=0\;,
\end{equation}
\begin{equation}\label{eom einstein ij em}
    \frac{2a''}{a} +\frac{a'^{2}}{a^{2}}-\frac{1}{a^{2}}-\frac{\kappa Q^{2}}{3a^{4}}-\frac{1}{L^{2}}=0\;,
\end{equation}
\begin{equation}\label{EOM chunEM}
    \nabla_{\mu}F^{\mu\nu}=0\;.
\end{equation}
We still set $a'=0,\;\tau=0$ in (\ref{eom einstein 00 em}) , we get
\begin{equation}
    a^{4}+a^{2}-\tilde{Q}^{2}=0,\quad\Delta=1+4\tilde{Q}^{2}>0\;.
\end{equation}
At this point, \(a\) has only one solution
\begin{equation}
    a^{2}=\frac{-1+\sqrt{1+4\tilde{Q}^{2}}}{2}\;.
\end{equation}
We continue to use equation (\ref{eom einstein ij em}) to study
its extremal conditions. By substituting \(a' = 0\), we get
\begin{equation}
    a''=\frac{a}{2}\left(\frac{1}{a^{2}}+\frac{\kappa Q^{2}}{3a^{4}}+\frac{1}{L^{2}}\right)>0\;.
\end{equation}

Therefore, in the case of a pure electromagnetic field with an
EAdS boundary, it is similar to the situation with the
$\frac{Q^{2}}{a^{8}}$ term, where the initial state of universe
evolving through the Euclidean wormhole gives rise to a collapsing
universe, as seen in section \ref{app a8}.




\end{document}